\documentclass[sigconf]{acmart}
\AtBeginDocument{%
  }

\setcopyright{acmlicensed}
\copyrightyear{2018}
\acmYear{2018}
\acmDOI{XXXXXXX.XXXXXXX}
\acmISBN{978-1-4503-XXXX-X/2018/06}

\begin{document}

\title{Glide-in-Place: Foot-Steered Differential-Drive for Hands-Free VR Locomotion}
\author{Bin Hu}
\authornote{These authors contributed equally to this work.}
\affiliation{%
  \institution{Tsinghua University}
  \country{China}}
\email{hub25@mails.tsinghua.edu.cn}
\orcid{0009-0001-7676-3130}

\author{Yang Liu}
\authornotemark[1]
\affiliation{%
 \institution{Beihang University}
 \country{China}}
\email{yangliu0238@gmail.com}
\orcid{0009-0005-7160-0614}

\author{Xizi Liu}
\affiliation{%
 \institution{Beihang University}
 \country{China}}
\email{Liuxizi@buaa.edu.cn}
\orcid{0009-0005-5600-7871}

\author{Qinggerou Xiao}
\affiliation{%
 \institution{Beihang University}
 \country{China}}
\email{grow1026@gmail.com}
\orcid{0009-0002-8581-0020}

\author{Xiru Wang}
\affiliation{%
 \institution{Beihang University}
 \country{China}}
\email{wangxiru@buaa.edu.cn}
\orcid{0009-0007-4074-0790}

\author{Zhe Yuan}
\affiliation{%
 \institution{Beihang University}
 \country{China}}
\email{22261062@buaa.edu.cn}

\author{Wen Ku}
\affiliation{%
 \institution{Tsinghua University}
 \country{China}}
\email{kw23@mails.tsinghua.edu.cn}

\author{Xiu Li}
\authornote{corresponding author.}
\affiliation{%
  \institution{Tsinghua University}
  \country{China}}
\email{li.xiu@sz.tsinghua.edu.cn}

\author{Yun Wang}
\orcid{0000-0001-9847-2636}
\authornotemark[2]
\affiliation{%
  \institution{Beihang University}
  \country{China}}
\email{wang_yun@buaa.edu.cn}

\renewcommand{\shortauthors}{Trovato et al.}


\begin{abstract}
Seated VR locomotion in constrained environments---homes,
offices, transit settings---calls for hardware that is
lightweight and deployable, steering that remains continuous
enough for curved motion, and a control channel that leaves
the hands free for concurrent interaction.
Inspired by the steering logic of self-balancing scooters, we
present \textit{Glide-in-Place}, a seated-foot locomotion
system that maps per-foot fore--aft pressure to a
differential-drive model: the two feet act as virtual wheels
whose relative drive continuously determines translation and
yaw. This lets users move forward, rotate in place, and follow
arcs in one unified vocabulary---without hand-held input or
discrete mode switches. We evaluated Glide-in-Place in a
counterbalanced within-subject study with 16 participants
against two baselines: joystick control and a seated
walking-in-place technique with discrete snap motions. Across
two steering-heavy navigation tasks---zig-zag path following
with multitasking and curved-path traversal---Glide-in-Place
was consistently faster than Seated-WIP,
reduced physical demand, and lowered fatigue-related discomfort
without significantly differing from joystick control on total
VRSQ. We position Glide-in-Place as a deployable
hardware-control design point for constrained seated VR: thin
insole sensing, continuous foot steering, and lightweight
calibration packaged in one compact artifact.
\end{abstract}


  
\begin{teaserfigure}
  \includegraphics[width=\textwidth]{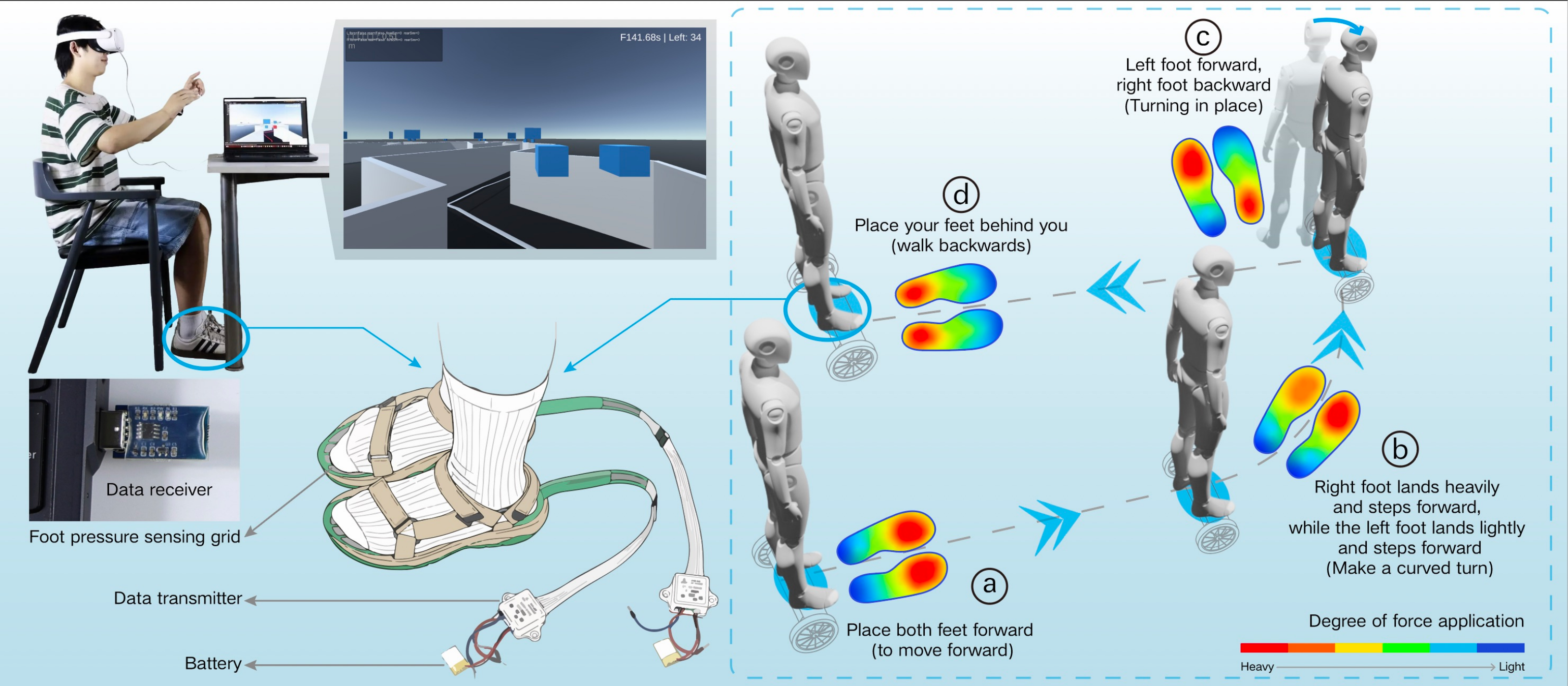}
  \caption{An overview of the Glide-in-Place locomotion technique.  (a) Shifting pressure forward on both feet moves the avatar forward. (b) Applying differential pressure---here, more on the right foot---initiates a curved turn (arc motion). (c) Opposing pressure (one foot forward, one back) results in an in-place turn. (d) Shifting pressure to the rear moves the avatar backward. The color gradient indicates the degree of force application.}
  \Description{}
  \label{fig:teaser}
\end{teaserfigure}


\maketitle

\section{Introduction}
\label{sec:intro}

VR locomotion remains difficult because no single technique
simultaneously solves space constraints, controllability,
embodiment, and comfort. In practice, many real sessions happen
in constrained seated settings---desks, offices, transit
environments, small rooms---where room-scale walking is
unavailable and users need their hands free for pointing,
selecting, and other concurrent
tasks~\cite{boletsis2022typology,prithul2021teleport_review,
zielasko2021sitOrNot}.

Existing options each give something up. Thumbsticks are
efficient but tie up the hands. Teleportation resists sickness
but fragments trajectories and weakens continuous
steering~\cite{bozgeyikli2016point,prithul2021teleport_review}.
Body-centric techniques such as walking-in-place and leaning
increase embodiment, but many require repeated stepping, torso
motion, or discrete state changes that fit constrained seated
use poorly~\cite{slater1995takingSteps,usoh1999walking,
ruddle2009benefits,wang2012leaningRevisited,
buttussi2021locomotionInPlace}. Dedicated lower-body hardware
can improve bodily engagement at the cost of deployability:
prior systems have used treadmill platforms, chair-based rigs,
or attachable wheeled footwear that are impractical as
lightweight everyday
interfaces~\cite{cyberithSiggraph,nguyenvo2021naviboardNavichair}. Seated-WIP brings foot-only
locomotion to stationary-chair VR, but its snap moves and snap
turns still decompose curved navigation into
rotate--then--translate segments, making smooth arc following
difficult~\cite{chan2024seatedWIP}. The gap is a seated
interface that steers continuously, keeps the hands free, and
deploys without specialized hardware.

Inspired by the steering logic of self-balancing scooters, we
present \textit{Glide-in-Place} (GIP): a deployable hardware design for continuous seated foot steering in VR. GIP maps
fore--aft pressure on each foot to a virtual wheel in a
differential-drive model, so the two feet together determine
translation and yaw continuously---without discrete mode
switches or hand-held input. The goal is simple: let the feet
steer and let the hands work.

\section{Related Work}
\label{sec:related}

\paragraph{Continuous versus discrete locomotion.}
VR locomotion techniques span a well-studied tradeoff between
continuous and discrete control. Continuous motion preserves
trajectory flow and embodiment, but comfort depends strongly
on how predictable and smooth the mapping
feels~\cite{buttussi2021locomotionInPlace,farmani2020discrete}.
Discrete techniques such as teleportation reduce optic-flow
conflict at the cost of trajectory
continuity~\cite{bozgeyikli2016point,prithul2021teleport_review}.
Body-centric interfaces---walking-in-place, leaning,
chair-based motion cueing---offer stronger bodily coupling than
thumbsticks, yet each commits to a different body segment and
interferes differently with manipulation or sustained seated
use~\cite{slater1995takingSteps,usoh1999walking,
ruddle2009benefits,wang2012leaningRevisited,
buttussi2021locomotionInPlace}.
None of these techniques simultaneously provides continuous
heading control, hands-free input, and insole-level
deployability in a seated setting.

\paragraph{Seated and foot-driven locomotion.}
Seated-WIP is the closest prior work: it demonstrates that
foot-only input can drive locomotion from a stationary chair,
but its quantized snap turns force users to decompose curved
paths into alternating rotate and translate
segments~\cite{chan2024seatedWIP}. More broadly, foot-driven
systems have largely inferred intent from step cycles or
gesture classifiers rather than exposing a continuous steering
channel~\cite{templeman1999virtualLocomotion,Feasel2008,
park2018wipImu,lee2019walkInPlace,sari2019vrfit,
kim2021userDefined}.
GIP addresses this gap by mapping pressure directly to a
continuous kinematic output, bypassing the
gesture-classification step entirely.

\paragraph{Hardware substrate and deployability.}
Treadmill platforms and chair-based rigs offer strong physical
fidelity but require dedicated space or infrastructure that
conflicts with everyday seated
use~\cite{cyberithSiggraph,nguyenvo2021naviboardNavichair}. Instrumented insoles occupy a
different point: they embed in existing footwear, preserve the
same sensing modality across seated and standing postures, and
expose continuous load transfer even when leg motion remains
subtle~\cite{chen2021plantarReview,bektas2021embodiedControl}.

GIP sits at the intersection of these three lines: it applies
a continuous differential-drive mapping to insole-based foot
pressure, targeting the deployable end of the hardware spectrum
while recovering the steering continuity that discrete
seated-foot baselines lose on curved paths.

\section{Glide-in-Place System Design}
\label{sec:system}
\paragraph{Design requirements.}
The system was guided by four target
setting. Locomotion remains hands-free so that users
could manipulate content while moving rather than treating
locomotion as a separate activity. It had to support continuous heading correction, because discrete seated-foot locomotion quantizes both rotation and
translation into fixed increments, forcing curved paths to
decompose into alternating rotate and translate steps rather
than allowing simultaneous smooth arc motion. The hardware
had to stay portable and discreet enough for desks, homes,
offices, and temporary setups rather than depending on chair
modification or room-scale infrastructure. Finally, the
control signal had to favor smooth pressure transfer over
impact-heavy stepping, because repeated foot lifting is both
noisier and less comfortable in constrained seated use.

\paragraph{Hardware.}
GIP uses a pair of full-length instrumented insoles, each
with 18 piezoresistive sensing elements aggregated into
forefoot and rearfoot regions, yielding four channels
overall: left/right forefoot and left/right rearfoot. The
insoles are 1\,mm thick and weigh about 20\,g each. Signals
are streamed from ankle-mounted microcontrollers at a minimum
rate of 100\,Hz over Bluetooth to a host PC and then
forwarded to Unity via TCP. End-to-end latency was
approximately 20\,ms. This hardware stack keeps the interface
shoe-integrated, portable, and deployable without chair
modification, room-scale sensing, or bulky foot
hardware~\cite{chen2021plantarReview,
nguyenvo2021naviboardNavichair}.

\begin{figure}[t]
    \centering
    \includegraphics[width=\linewidth]{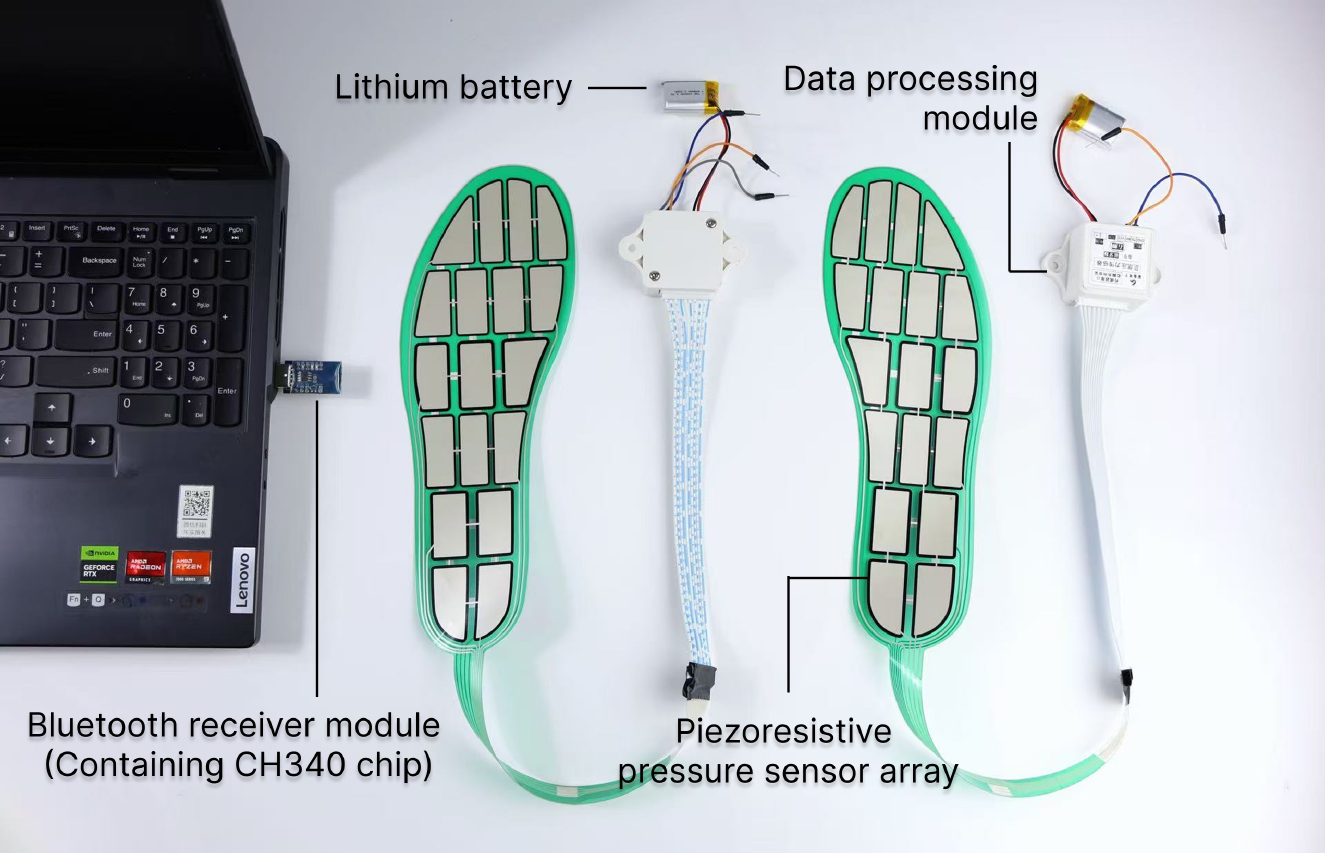}
    \caption{Hardware path. Each foot uses a pressure-sensing
    insole, and an ankle-mounted microcontroller streams four
    aggregated channels to the host PC.}
    \Description{A hardware diagram showing pressure-sensing
    insoles, ankle-mounted electronics, and the data path to
    the host PC and Unity runtime.}
    \label{fig:HardWare}
\end{figure}

\paragraph{Signal processing.}
The control pipeline is organized around neutral calibration,
drift suppression, and continuous kinematic mapping. At the
start of each block, users hold a neutral posture for
1--2\,s so that per-zone baselines can be estimated and
subtracted online. The calibrated fore--aft pressure signals
are then filtered with hysteretic dead-zones, exponential
smoothing, and a short minimum-hold debounce. If both feet
remain neutral for a sustained interval, the baseline is
automatically re-estimated to absorb slow drift. These steps
were chosen to favor predictability over peak responsiveness:
seated locomotion tolerates a modest speed penalty more
readily than it tolerates jitter or accidental yaw. This
tuning choice is later reflected in the lower fatigue-related
discomfort scores relative to Seated-WIP, where abrupt
snap-triggered motion imposed a less predictable control
response.

\paragraph{Differential-drive mapping.}
The key design choice is the differential-drive map, adapted
from the steering metaphor of self-balancing scooters. These
three interaction behaviors---forward travel, in-place
turning, and arc motion---emerge naturally from a single
mapping. Let the filtered fore--aft commands from the left
and right feet be mapped to normalized wheel inputs
$u_L, u_R \in [-1,1]$, then scaled to virtual wheel speeds
$v_L$ and $v_R$:
\begin{equation}
u_L = f(s_L),\quad u_R = f(s_R),\quad
v_L = V_{\max}u_L,\quad v_R = V_{\max}u_R.
\end{equation}
Avatar motion then follows standard differential-drive
kinematics with virtual track width $W$:
\begin{equation}
v = \frac{v_R + v_L}{2}, \qquad
\omega = \frac{v_R - v_L}{W}.
\end{equation}
For unequal wheel speeds, the instantaneous turning radius
is:
\begin{equation}
R = \frac{v}{\omega} = \frac{W}{2}\cdot
\frac{v_R + v_L}{v_R - v_L}.
\end{equation}
If $v_L \approx v_R$, then $\omega \approx 0$ and the user
moves nearly straight. If $v_L \approx -v_R$, then
$v \approx 0$ and the avatar turns in place. If the two
wheel speeds share the same sign but differ in magnitude,
the avatar follows an arc, and increasing left--right
pressure imbalance tightens that arc by reducing $R$.

The intended contribution is not the mathematical novelty
of differential drive in isolation, but its packaging as a
deployable seated-foot interaction grammar that matches the
constraints of hands-free VR much better than
step-triggered control~\cite{templeman1999virtualLocomotion,
kim2021userDefined}.

\begin{figure}[t]
    \centering
    \includegraphics[width=\linewidth]{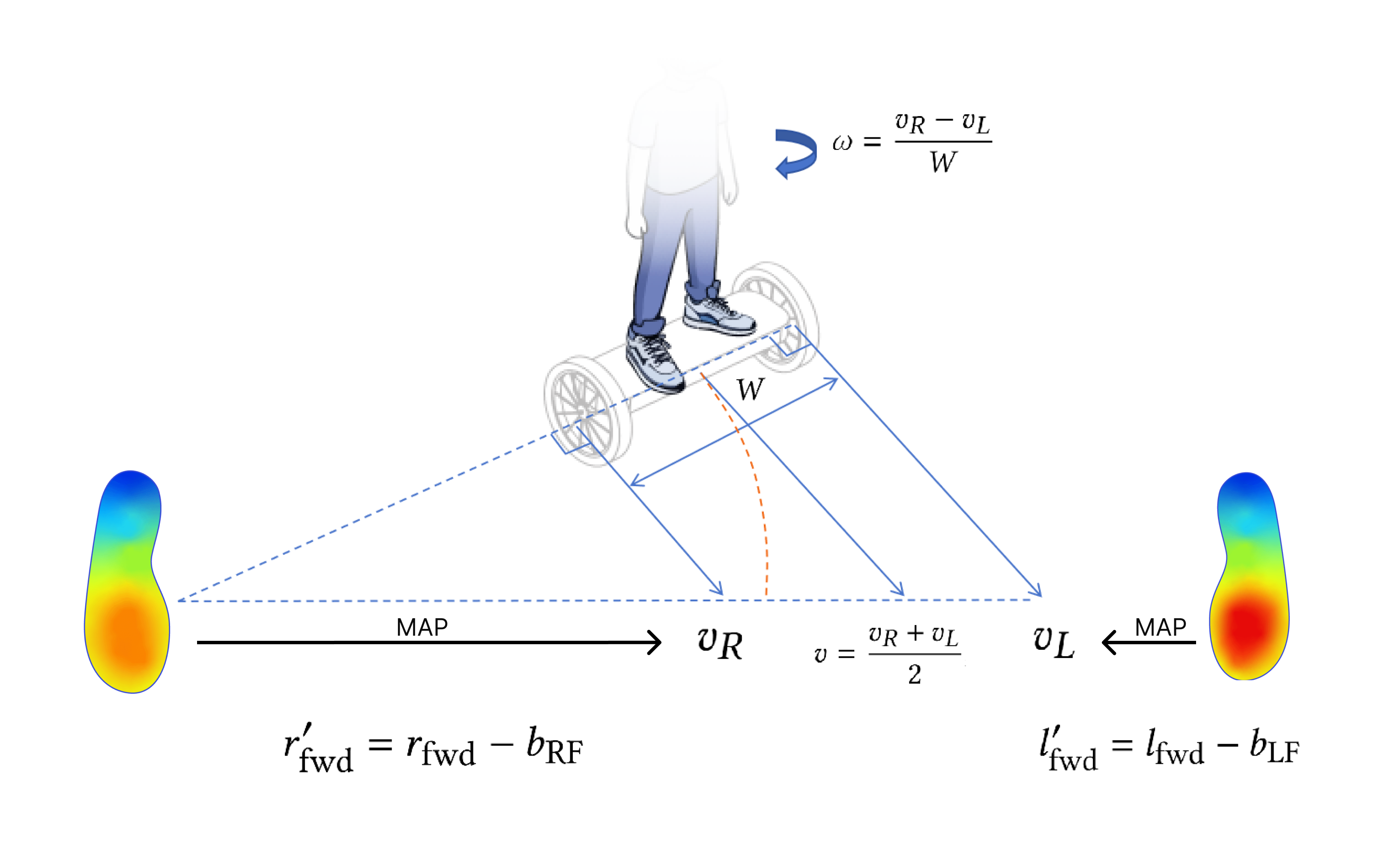}
    \caption{Differential-drive interpretation of GIP. The
    left and right feet act as virtual wheels whose relative
    drive determines translation, in-place turning, and arc
    curvature.}
    \Description{A differential-drive diagram showing two
    virtual wheels controlled by the left and right feet,
    together with the resulting avatar translation and
    rotation variables.}
    \label{fig:Math}
\end{figure}

\section{Evaluation}
\label{sec:userstudy}

We compared GIP against \textbf{Controller} (right-hand thumbstick) and \textbf{Seated-WIP} (chair-bound walking-in-place with discrete snap locomotion and snap turning~\cite{chan2024seatedWIP}) in a within-subject experiment across three task blocks of increasing steering demand.

\subsection{Study Design}

Each participant completed all three conditions in all three task blocks. Blocks were fixed in order from open-space waypoint travel (Study~1) to zig-zag path following (Study~2) to curved-path traversal (Study~3). Condition order within each block was counterbalanced using a six-sequence Williams design, controlling ordinal position and first-order carryover. Each condition used a geometrically matched but distinct layout; layout-to-condition mapping was rotated across groups.

\subsection{Apparatus and Participants}

All environments ran on a Meta Quest 2 in Unity. Seated-WIP and GIP were foot-driven, leaving the hands free for tasks. Participants sat on a fixed-height, non-swivel chair with a footrest. GIP plantar-pressure signals were captured via thin insole prototypes and streamed to the runtime in real time. The Seated-WIP baseline was implemented following Chan et al.~\cite{chan2024seatedWIP}, preserving the original snap step distance and snap turn angle, adapted to trigger on plantar-pressure transients.

Sixteen participants (8 female, 8 male; age 19--25, $M = 21.8$, $SD = 1.8$) were recruited from the university community, all of them have VR experience. All reported normal or corrected-to-normal vision and no vestibular disorders, gave informed consent, and received compensation. The study was approved by the authors' Institutional Review Board. Sample size was close toprior seated-foot locomotion studies of similar within-subject designs (Seated-WIP~\cite{chan2024seatedWIP}, $N=18$).

\subsection{Procedure and Tasks}
Each session began with a technique walkthrough and per-condition tutorial. One untimed practice trial preceded every task-condition block. Conditions were separated by 2-minute rests; task blocks by 5-minute breaks. GIP was recalibrated before each block and after each break.
\begin{figure}[htbp]
    \centering
    \includegraphics[width=\linewidth]{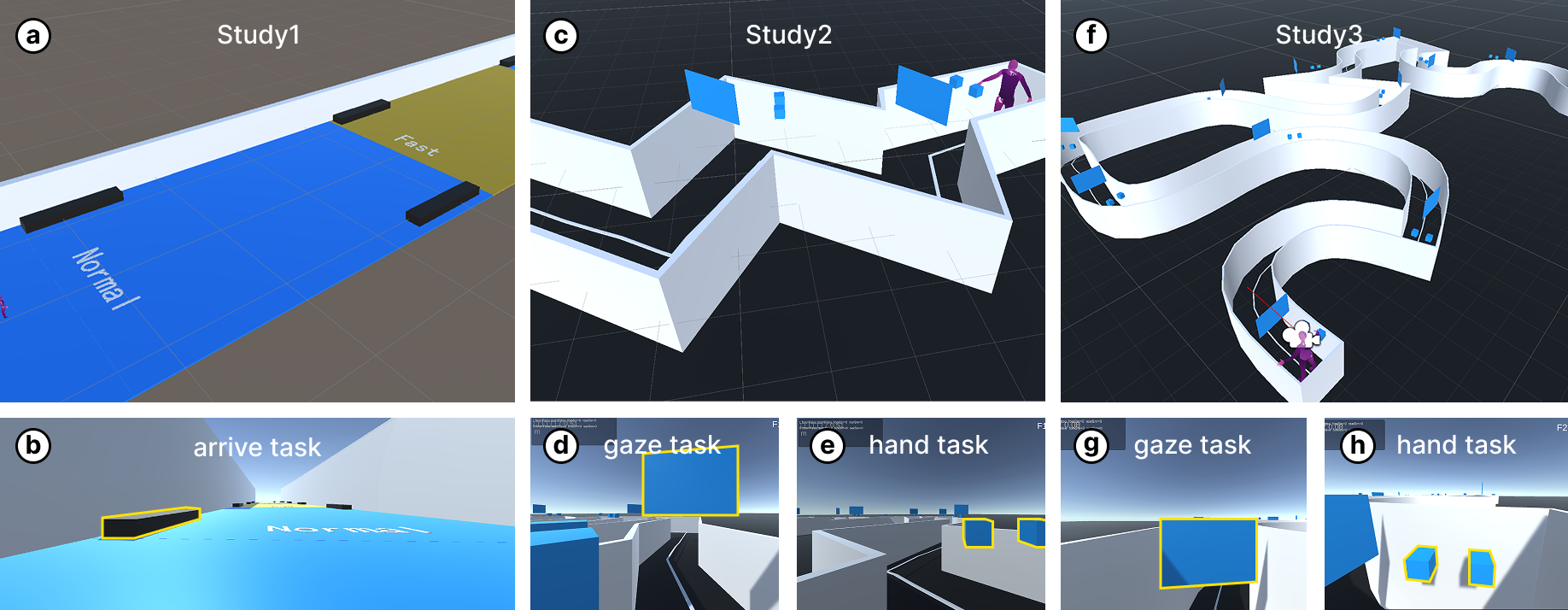}
    \caption{Evaluation tasks. Study~1 measured open-space waypoint travel, whereas Study~2 and Study~3 stressed structured steering and multitasking through zig-zag and curved-path navigation.}
    \Description{A composite figure showing the three evaluation tasks: open-space waypoint travel, zig-zag path following with gaze and hand subtasks, and curved-path traversal with the same multitasking demands.}
    \label{fig:Study}
\end{figure}
Study~1 was an open-space arrival task in a 160\,m environment with no prescribed route. Study~2 required following a zig-zag polyline while responding to concurrent gaze and hand prompts; frequent heading changes exposed the cost of Seated-WIP's rotate-then-translate pattern. Study~3 replaced the polyline with circular arcs under the same multitasking demands---the most diagnostic test for GIP, as arc traversal requires simultaneous, continuous heading and translation control. After each block, participants completed NASA-TLX~\cite{hart1988nasaTLX} and VRSQ~\cite{kim2018vrsq} and offered brief verbal comments; at session end they ranked and justified their technique preferences.




\subsection{Measures and Analysis}
 
The primary outcome was completion time across all three task blocks. Study~1 (open-space arrival) involved locomotion only; Studies~2 and~3 added concurrent gaze and hand subtasks, and subtask completion was additionally logged to support interpretation of the hands-free benefit. NASA-TLX and VRSQ were collected after every task-condition block to characterize workload and perceptual discomfort.
 
Locomotion condition was the primary within-subject factor, analyzed separately per task block. Ordinal position, counterbalancing group, path instance, and trial index were included as nuisance covariates; arc radius was additionally controlled in Study~3. Participant was modeled as a random intercept. Pairwise comparisons used Holm correction; effect sizes are reported as $d_z$.

\subsection{Results}
 \begin{figure*}[htbp]
    \centering
    \includegraphics[width=\linewidth]{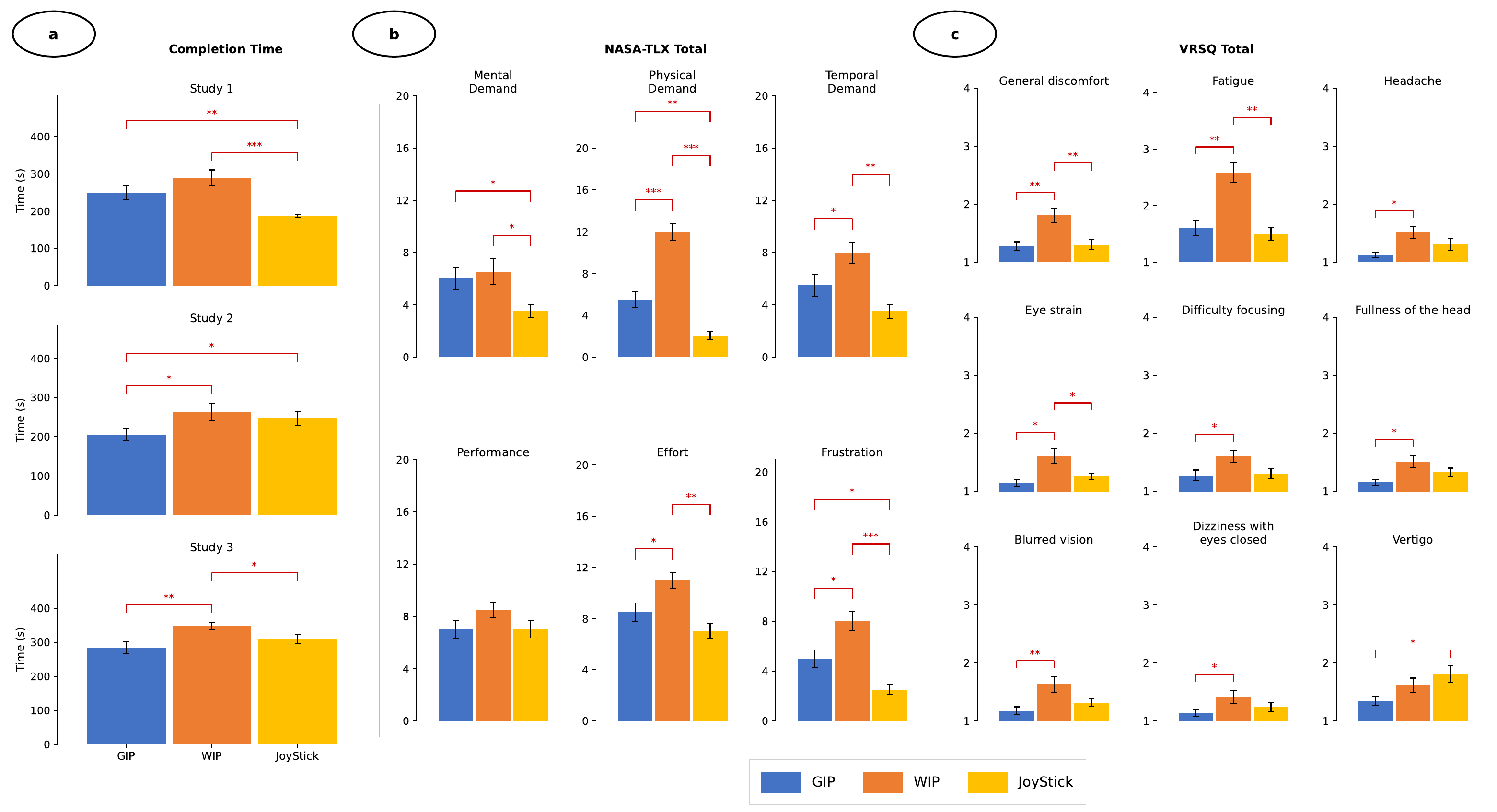}
    \caption{Quantitative results across all three conditions (GIP, WIP, JoyStick). 
(a)~Completion time per task block: GIP was fastest in Studies~2 and~3, with 
the advantage over WIP widening as steering demand increased; Study~1 showed no 
reliable difference between foot-driven techniques. 
(b)~NASA-TLX subscale ratings: WIP incurred substantially higher physical demand, 
effort, and frustration than both GIP and JoyStick across all dimensions. 
(c)~VRSQ symptom ratings: WIP was rated highest on fatigue and most perceptual 
discomfort items; GIP matched JoyStick on nearly every symptom. 
Error bars denote $\pm$1 SE; brackets indicate Holm-corrected pairwise differences 
($^{*}p<.05$, $^{**}p<.01$, $^{***}p<.001$).}

    \label{fig:results}
\end{figure*}
\subsubsection{Completion Time}
 
GIP was faster than Seated-WIP across all three blocks, with the margin widening as steering demand increased (Figure~\ref{fig:results}a).
 
In Study~1, GIP ($M = 249$\,s, $SD = 81$) was numerically faster than Seated-WIP ($M = 290$\,s, $SD = 91$), but the difference did not survive Holm correction ($p = .323$, $d_z = 0.32$). Controller was fastest ($M = 188$\,s, $SD = 18$), differing from both GIP ($p < .01$, $d_z = 0.77$) and Seated-WIP ($p < .001$, $d_z = 1.26$). The null result is expected: open-space travel imposes no steering continuity requirement, so Seated-WIP's snap-turn overhead seldom compounds.
 
Study~2 produced the first reliable separation between foot-driven techniques. GIP ($M = 206$\,s, $SD = 74$) was faster than both Seated-WIP ($M = 264$\,s, $SD = 69$; $p < .05$, $d_z = 0.56$) and Controller ($M = 247$\,s, $SD = 91$; $p < .05$, $d_z = 0.66$); Controller and Seated-WIP did not differ ($p = .298$). The advantage emerged because each heading change forced Seated-WIP into a discrete rotate-then-translate cycle.
 
The gap widened in Study~3: GIP ($M = 285$\,s, $SD = 90$) outpaced Seated-WIP ($M = 349$\,s, $SD = 40$; $p < .01$, $d_z = 0.88$); Seated-WIP was also slower than Controller ($M = 310$\,s, $SD = 58$; $p < .05$, $d_z = 0.70$). Arc traversal left no opportunity to chain straight segments, exposing the full cost of quantized heading control.
 
\subsubsection{Workload (NASA-TLX)}
 
Physical demand, effort, frustration, and temporal demand all placed Seated-WIP above GIP (Figure~\ref{fig:results}b). Physical demand was highest for Seated-WIP ($M = 12.0$), well above GIP ($M = 5.5$; $p < .001$, $d_z = 1.47$) and Controller ($M = 2.1$; $p < .001$, $d_z = 2.77$). Effort (Seated-WIP $M = 11.0$ vs.\ GIP $M = 8.5$; $p < .05$, $d_z = 0.61$), frustration ($M = 8.0$ vs.\ $5.0$; $p < .05$, $d_z = 0.65$), and temporal demand ($M = 8.0$ vs.\ $5.5$; $p < .05$, $d_z = 0.50$) replicated this pattern. Mental demand and performance did not differ reliably between GIP and Seated-WIP ($p > .10$), indicating that GIP's physical advantage did not extend to perceived cognitive load.
 
\subsubsection{Discomfort (VRSQ)}
 
GIP rated below Seated-WIP on nearly every VRSQ dimension and matched Controller throughout (Figure~\ref{fig:results}c). Fatigue showed the largest effect: Seated-WIP ($M = 2.58$) exceeded both GIP ($M = 1.61$; $p < .01$, $d_z = 1.01$) and Controller ($M = 1.50$; $p < .01$, $d_z = 1.19$), with GIP and Controller indistinguishable ($p = .717$). General discomfort ($d_z = 0.83$) and five further symptoms---blurred vision, eye strain, difficulty focusing, headache, fullness of the head---all showed reliable GIP--Seated-WIP differences ($p < .05$, $d_z = 0.69$--$0.93$) with non-significant GIP--Controller comparisons. Vertigo was the sole exception ($p > .12$ across all pairs), likely reflecting individual variability in susceptibility to rotational vection.
 
\subsubsection{Qualitative Results}
 
Post-session interviews and verbal comments converged on three themes.
 
\paragraph{Steering continuity.}
The most common complaint about Seated-WIP was its quantized heading control. P7 described having to ``stop, snap, then go again'' on every curve, while P11 said the technique ``never let you just lean into a turn.'' GIP was contrasted as feeling closer to ``holding a curve'' (P15)---participants reported adjusting heading without interrupting forward motion, which they linked directly to lower mental effort on the arc task. P3 summarized the difference as ``WIP makes you plan every turn; GIP just goes where you push.''
 
\paragraph{Physical effort and multitasking.}
Seated-WIP's discrete stepping was described as progressively tiring across longer trials. P9 noted that ``by the third arc I was thinking more about my feet than the target,'' an observation consistent with the elevated frustration and temporal demand ratings. GIP's pressure-based input was perceived as more passive: P5 said it ``felt like resting your foot, not doing exercise.'' Both foot-driven conditions were credited with freeing the hands; Controller users typically became aware of the hand-occupancy cost only once a trial required simultaneous pointing or selecting, with P12 remarking that ``you don't notice how busy your thumb is until you need your hand for something else.''
 
\paragraph{Preference.}
End-of-session rankings aligned with the quantitative results. Controller was preferred for open-space travel due to its low learning overhead. GIP was ranked highest on the zig-zag and arc tasks by the majority of participants, with several citing the combination of continuous steering and free hands as the deciding factor. Seated-WIP was ranked lowest on structured-path trials; P4 captured the shared sentiment: ``it gets the job done, but it wears you out and you're always a step behind the path.''

\section{Discussion}
\label{sec:discussion}

GIP occupies a part of the design space where room-scale walking is impossible, joystick control is interactionally costly, and discrete seated-foot stepping breaks down on curves. Its contribution is the package: thin insole sensing, lightweight calibration, and a differential-drive interaction grammar combined into a deployable seated-foot interface. In our 16-participant counterbalanced study, GIP consistently outperformed the nearest discrete seated-foot baseline on steering-heavy tasks while preserving the hands-free benefit that motivates lower-body input in the first place.

Controller's NASA-TLX advantage was concentrated in mental demand, physical demand, and frustration---dimensions where familiarity with the default VR input reduces perceived cost regardless of task structure. Temporal demand, effort, and performance did not differ reliably between GIP and Controller, suggesting that GIP closes the workload gap on dimensions driven by steering continuity rather than input convention. The relevant question is therefore not whether GIP matches Controller overall, but whether it recovers enough steering quality to justify moving locomotion off the hands---and in structured navigation with concurrent interaction, our results suggest it does.



\section{Limitations and Future Work}
\label{sec:limitations}
The current prototype uses a single insole size and does not adapt to foot morphology or footwear stiffness. While aggregating 18 sensors into four channels robustly accommodates varying foot placements, future versions could simplify the hardware to four low-cost discrete sensors. Furthermore, extended use may introduce one-foot bias, calibration drift (partially mitigated by automatic recentering), and localized lower-leg fatigue from sustained pressure exertion. These constraints represent the immediate barriers to broader deployment.

The most direct next steps are lightweight personalization---per-user gain balancing, footwear-aware normalization, and invisible drift-recovery strategies---alongside evaluations across diverse shoe types and longer durations. Beyond hardware refinement, future work should examine whether GIP's continuous steering advantage extends to dynamic environments with obstacles, non-seated postures, or tasks requiring higher-precision spatial control than the navigation scenarios tested here.

\section{Conclusion}
\label{sec:conclusion}

Glide-in-Place reframes seated VR locomotion as a deployable hardware-control problem. By combining thin insole sensing with a differential-drive mapping, it exposes continuous foot steering in a format that fits constrained seated use better than step-triggered baselines. In a 16-participant counterbalanced study, GIP consistently improved over Seated-WIP on completion time, workload, and discomfort across steering-heavy tasks, while Controller remained fastest overall. These results position GIP as a concrete design point for continuous seated foot steering in constrained VR contexts rather than a universal locomotion method.

\clearpage 
\bibliographystyle{ACM-Reference-Format}
\bibliography{main}


\begin{thebibliography}{23}


\ifx \showCODEN    \undefined \def \showCODEN     #1{\unskip}     \fi
\ifx \showISBNx    \undefined \def \showISBNx     #1{\unskip}     \fi
\ifx \showISBNxiii \undefined \def \showISBNxiii  #1{\unskip}     \fi
\ifx \showISSN     \undefined \def \showISSN      #1{\unskip}     \fi
\ifx \showLCCN     \undefined \def \showLCCN      #1{\unskip}     \fi
\ifx \shownote     \undefined \def \shownote      #1{#1}          \fi
\ifx \showarticletitle \undefined \def \showarticletitle #1{#1}   \fi
\ifx \showURL      \undefined \def \showURL       {\relax}        \fi
\providecommand\bibfield[2]{#2}
\providecommand\bibinfo[2]{#2}
\providecommand\natexlab[1]{#1}
\providecommand\showeprint[2][]{arXiv:#2}

\bibitem[Bekta{\c{s}} et~al\mbox{.}(2021)]%
        {bektas2021embodiedControl}
\bibfield{author}{\bibinfo{person}{Kenan Bekta{\c{s}}}, \bibinfo{person}{Tyler Thrash}, \bibinfo{person}{Mark~A. van Raai}, \bibinfo{person}{Patrik K{\"u}nzler}, {and} \bibinfo{person}{Richard Hahnloser}.} \bibinfo{year}{2021}\natexlab{}.
\newblock \showarticletitle{The systematic evaluation of an embodied control interface for virtual reality}.
\newblock \bibinfo{journal}{\emph{PLOS ONE}} \bibinfo{volume}{16}, \bibinfo{number}{12} (\bibinfo{year}{2021}), \bibinfo{pages}{e0259977}.
\newblock
\href{https://doi.org/10.1371/journal.pone.0259977}{doi:\nolinkurl{10.1371/journal.pone.0259977}}


\bibitem[Boletsis and Chasanidou(2022)]%
        {boletsis2022typology}
\bibfield{author}{\bibinfo{person}{Costas Boletsis} {and} \bibinfo{person}{Dimitra Chasanidou}.} \bibinfo{year}{2022}\natexlab{}.
\newblock \showarticletitle{A Typology of Virtual Reality Locomotion Techniques}.
\newblock \bibinfo{journal}{\emph{Multimodal Technologies and Interaction}} \bibinfo{volume}{6}, \bibinfo{number}{9} (\bibinfo{year}{2022}), \bibinfo{pages}{72}.
\newblock
\href{https://doi.org/10.3390/mti6090072}{doi:\nolinkurl{10.3390/mti6090072}}


\bibitem[Bozgeyikli et~al\mbox{.}(2016)]%
        {bozgeyikli2016point}
\bibfield{author}{\bibinfo{person}{Evren Bozgeyikli}, \bibinfo{person}{Andrew Raij}, \bibinfo{person}{Srinivas Katkoori}, {and} \bibinfo{person}{Rajiv Dubey}.} \bibinfo{year}{2016}\natexlab{}.
\newblock \showarticletitle{Point \& Teleport Locomotion Technique for Virtual Reality}. In \bibinfo{booktitle}{\emph{Proceedings of the 2016 Annual Symposium on Computer-Human Interaction in Play}}. \bibinfo{publisher}{Association for Computing Machinery}, \bibinfo{address}{New York, NY, USA}, \bibinfo{pages}{205--216}.
\newblock
\href{https://doi.org/10.1145/2967934.2968105}{doi:\nolinkurl{10.1145/2967934.2968105}}


\bibitem[Buttussi and Chittaro(2021)]%
        {buttussi2021locomotionInPlace}
\bibfield{author}{\bibinfo{person}{Fabio Buttussi} {and} \bibinfo{person}{Luca Chittaro}.} \bibinfo{year}{2021}\natexlab{}.
\newblock \showarticletitle{Locomotion in Place in {Virtual Reality}: A Comparative Evaluation of Joystick, Teleport, and Leaning}.
\newblock \bibinfo{journal}{\emph{IEEE Transactions on Visualization and Computer Graphics}} \bibinfo{volume}{27}, \bibinfo{number}{1} (\bibinfo{year}{2021}), \bibinfo{pages}{125--136}.
\newblock
\href{https://doi.org/10.1109/TVCG.2019.2928304}{doi:\nolinkurl{10.1109/TVCG.2019.2928304}}


\bibitem[Cakmak and Hager(2014)]%
        {cyberithSiggraph}
\bibfield{author}{\bibinfo{person}{Tuncay Cakmak} {and} \bibinfo{person}{Holger Hager}.} \bibinfo{year}{2014}\natexlab{}.
\newblock \showarticletitle{Cyberith Virtualizer: A Locomotion Device for Virtual Reality}. In \bibinfo{booktitle}{\emph{ACM SIGGRAPH 2014 Emerging Technologies}}. \bibinfo{publisher}{Association for Computing Machinery}, \bibinfo{address}{New York, NY, USA}, \bibinfo{pages}{1--1}.
\newblock
\href{https://doi.org/10.1145/2614066.2614105}{doi:\nolinkurl{10.1145/2614066.2614105}}


\bibitem[Chan et~al\mbox{.}(2024)]%
        {chan2024seatedWIP}
\bibfield{author}{\bibinfo{person}{Liwei Chan}, \bibinfo{person}{Tzu-Wei Mi}, \bibinfo{person}{Zhung~Hao Hsueh}, \bibinfo{person}{Yi-Ci Huang}, {and} \bibinfo{person}{Ming~Yun Hsu}.} \bibinfo{year}{2024}\natexlab{}.
\newblock \showarticletitle{Seated-WIP: Enabling Walking-in-Place Locomotion for Stationary Chairs in Confined Spaces}. In \bibinfo{booktitle}{\emph{Proceedings of the CHI Conference on Human Factors in Computing Systems}}. \bibinfo{publisher}{Association for Computing Machinery}, \bibinfo{address}{New York, NY, USA}, \bibinfo{pages}{1--13}.
\newblock
\href{https://doi.org/10.1145/3613904.3642395}{doi:\nolinkurl{10.1145/3613904.3642395}}


\bibitem[Chen et~al\mbox{.}(2022)]%
        {chen2021plantarReview}
\bibfield{author}{\bibinfo{person}{Jun-Liang Chen}, \bibinfo{person}{Yan-Ning Dai}, \bibinfo{person}{Nicolas~S. Grimaldi}, \bibinfo{person}{Jing-Jing Lin}, \bibinfo{person}{Bo-Yi Hu}, \bibinfo{person}{Yin-Feng Wu}, {and} \bibinfo{person}{Shuo Gao}.} \bibinfo{year}{2022}\natexlab{}.
\newblock \showarticletitle{Plantar Pressure-Based Insole Gait Monitoring Techniques for Diseases Monitoring and Analysis: {A} Review}.
\newblock \bibinfo{journal}{\emph{Advanced Materials Technologies}} \bibinfo{volume}{7}, \bibinfo{number}{1} (\bibinfo{year}{2022}), \bibinfo{pages}{2100566}.
\newblock
\href{https://doi.org/10.1002/admt.202100566}{doi:\nolinkurl{10.1002/admt.202100566}}


\bibitem[Farmani and Teather(2020)]%
        {farmani2020discrete}
\bibfield{author}{\bibinfo{person}{Yasin Farmani} {and} \bibinfo{person}{Robert~J. Teather}.} \bibinfo{year}{2020}\natexlab{}.
\newblock \showarticletitle{Evaluating discrete viewpoint control to reduce cybersickness in virtual reality}.
\newblock \bibinfo{journal}{\emph{Virtual Reality}} \bibinfo{volume}{24}, \bibinfo{number}{4} (\bibinfo{year}{2020}), \bibinfo{pages}{645--664}.
\newblock
\href{https://doi.org/10.1007/s10055-020-00425-x}{doi:\nolinkurl{10.1007/s10055-020-00425-x}}


\bibitem[Feasel et~al\mbox{.}(2008)]%
        {Feasel2008}
\bibfield{author}{\bibinfo{person}{Jeff Feasel}, \bibinfo{person}{Mary~C. Whitton}, {and} \bibinfo{person}{Jeremy~D. Wendt}.} \bibinfo{year}{2008}\natexlab{}.
\newblock \showarticletitle{LLCM-WIP: Low-Latency, Continuous-Motion Walking-in-Place}. In \bibinfo{booktitle}{\emph{2008 IEEE Symposium on 3D User Interfaces}}. \bibinfo{publisher}{IEEE}, \bibinfo{address}{Piscataway, NJ, USA}, \bibinfo{pages}{97--104}.
\newblock
\href{https://doi.org/10.1109/3DUI.2008.4476598}{doi:\nolinkurl{10.1109/3DUI.2008.4476598}}


\bibitem[Hart and Staveland(1988)]%
        {hart1988nasaTLX}
\bibfield{author}{\bibinfo{person}{Sandra~G. Hart} {and} \bibinfo{person}{Lowell~E. Staveland}.} \bibinfo{year}{1988}\natexlab{}.
\newblock \showarticletitle{Development of {NASA-TLX} ({Task Load Index}): Results of Empirical and Theoretical Research}.
\newblock In \bibinfo{booktitle}{\emph{Human Mental Workload}}, \bibfield{editor}{\bibinfo{person}{Peter~A. Hancock} {and} \bibinfo{person}{Najmedin Meshkati}} (Eds.). \bibinfo{publisher}{Elsevier}, \bibinfo{address}{Amsterdam, the Netherlands}, \bibinfo{pages}{139--183}.
\newblock
\href{https://doi.org/10.1016/S0166-4115(08)62386-9}{doi:\nolinkurl{10.1016/S0166-4115(08)62386-9}}


\bibitem[Kim et~al\mbox{.}(2018)]%
        {kim2018vrsq}
\bibfield{author}{\bibinfo{person}{Hyun~K. Kim}, \bibinfo{person}{Jaehyun Park}, \bibinfo{person}{Yeongcheol Choi}, {and} \bibinfo{person}{Mungyeong Choe}.} \bibinfo{year}{2018}\natexlab{}.
\newblock \showarticletitle{{Virtual Reality Sickness Questionnaire (VRSQ)}: Motion sickness measurement index in a virtual reality environment}.
\newblock \bibinfo{journal}{\emph{Applied Ergonomics}}  \bibinfo{volume}{69} (\bibinfo{year}{2018}), \bibinfo{pages}{66--73}.
\newblock
\href{https://doi.org/10.1016/j.apergo.2017.12.016}{doi:\nolinkurl{10.1016/j.apergo.2017.12.016}}


\bibitem[Kim and Xiong(2021)]%
        {kim2021userDefined}
\bibfield{author}{\bibinfo{person}{Woojoo Kim} {and} \bibinfo{person}{Shuping Xiong}.} \bibinfo{year}{2021}\natexlab{}.
\newblock \showarticletitle{User-defined walking-in-place gestures for {VR} locomotion}.
\newblock \bibinfo{journal}{\emph{International Journal of Human-Computer Studies}}  \bibinfo{volume}{152} (\bibinfo{year}{2021}), \bibinfo{pages}{102648}.
\newblock
\href{https://doi.org/10.1016/j.ijhcs.2021.102648}{doi:\nolinkurl{10.1016/j.ijhcs.2021.102648}}


\bibitem[Lee and Hwang(2019)]%
        {lee2019walkInPlace}
\bibfield{author}{\bibinfo{person}{Juyoung Lee} {and} \bibinfo{person}{Jae-In Hwang}.} \bibinfo{year}{2019}\natexlab{}.
\newblock \showarticletitle{Walk-in-Place Navigation in {VR}}. In \bibinfo{booktitle}{\emph{Proceedings of the 2019 ACM International Conference on Interactive Surfaces and Spaces}}. \bibinfo{publisher}{Association for Computing Machinery}, \bibinfo{address}{New York, NY, USA}, \bibinfo{pages}{427--430}.
\newblock
\href{https://doi.org/10.1145/3343055.3361926}{doi:\nolinkurl{10.1145/3343055.3361926}}


\bibitem[Nguyen-Vo et~al\mbox{.}(2021)]%
        {nguyenvo2021naviboardNavichair}
\bibfield{author}{\bibinfo{person}{Thinh Nguyen-Vo}, \bibinfo{person}{Bernhard~E. Riecke}, \bibinfo{person}{Wolfgang Stuerzlinger}, \bibinfo{person}{Duc-Minh Pham}, {and} \bibinfo{person}{Ernst Kruijff}.} \bibinfo{year}{2021}\natexlab{}.
\newblock \showarticletitle{{NaviBoard} and {NaviChair}: Limited Translation Combined with Full Rotation for Efficient Virtual Locomotion}.
\newblock \bibinfo{journal}{\emph{IEEE Transactions on Visualization and Computer Graphics}} \bibinfo{volume}{27}, \bibinfo{number}{1} (\bibinfo{year}{2021}), \bibinfo{pages}{165--177}.
\newblock
\href{https://doi.org/10.1109/TVCG.2019.2935730}{doi:\nolinkurl{10.1109/TVCG.2019.2935730}}


\bibitem[Park et~al\mbox{.}(2018)]%
        {park2018wipImu}
\bibfield{author}{\bibinfo{person}{Chanho Park}, \bibinfo{person}{Kyungho Jang}, {and} \bibinfo{person}{Junsuk Lee}.} \bibinfo{year}{2018}\natexlab{}.
\newblock \showarticletitle{Walking-in-Place for {VR} Navigation Independent of Gaze Direction Using a Waist-Worn Inertial Measurement Unit}. In \bibinfo{booktitle}{\emph{2018 IEEE International Symposium on Mixed and Augmented Reality Adjunct}}. \bibinfo{publisher}{IEEE}, \bibinfo{address}{Piscataway, NJ, USA}, \bibinfo{pages}{254--257}.
\newblock
\href{https://doi.org/10.1109/ISMAR-Adjunct.2018.00079}{doi:\nolinkurl{10.1109/ISMAR-Adjunct.2018.00079}}


\bibitem[Prithul et~al\mbox{.}(2021)]%
        {prithul2021teleport_review}
\bibfield{author}{\bibinfo{person}{Aniruddha Prithul}, \bibinfo{person}{Isayas~Berhe Adhanom}, {and} \bibinfo{person}{Eelke Folmer}.} \bibinfo{year}{2021}\natexlab{}.
\newblock \showarticletitle{Teleportation in Virtual Reality; A Mini-Review}.
\newblock \bibinfo{journal}{\emph{Frontiers in Virtual Reality}}  \bibinfo{volume}{2} (\bibinfo{year}{2021}), \bibinfo{pages}{730792}.
\newblock
\href{https://doi.org/10.3389/frvir.2021.730792}{doi:\nolinkurl{10.3389/frvir.2021.730792}}


\bibitem[Ruddle and Lessels(2009)]%
        {ruddle2009benefits}
\bibfield{author}{\bibinfo{person}{Roy~A. Ruddle} {and} \bibinfo{person}{Simon Lessels}.} \bibinfo{year}{2009}\natexlab{}.
\newblock \showarticletitle{The benefits of using a walking interface to navigate virtual environments}.
\newblock \bibinfo{journal}{\emph{ACM Transactions on Computer-Human Interaction}} \bibinfo{volume}{16}, \bibinfo{number}{1} (\bibinfo{year}{2009}), \bibinfo{pages}{1--18}.
\newblock
\href{https://doi.org/10.1145/1502800.1502805}{doi:\nolinkurl{10.1145/1502800.1502805}}


\bibitem[Sari and Kucukyilmaz(2019)]%
        {sari2019vrfit}
\bibfield{author}{\bibinfo{person}{Sercan Sari} {and} \bibinfo{person}{Ayse Kucukyilmaz}.} \bibinfo{year}{2019}\natexlab{}.
\newblock \showarticletitle{{VR-Fit}: Walking-in-Place Locomotion with Real Time Step Detection for {VR}-Enabled Exercise}.
\newblock In \bibinfo{booktitle}{\emph{Mobile Web and Intelligent Information Systems}}. \bibinfo{publisher}{Springer International Publishing}, \bibinfo{address}{Cham, Switzerland}, \bibinfo{pages}{255--266}.
\newblock
\href{https://doi.org/10.1007/978-3-030-27192-3_20}{doi:\nolinkurl{10.1007/978-3-030-27192-3_20}}


\bibitem[Slater et~al\mbox{.}(1995)]%
        {slater1995takingSteps}
\bibfield{author}{\bibinfo{person}{Mel Slater}, \bibinfo{person}{Martin Usoh}, {and} \bibinfo{person}{Anthony Steed}.} \bibinfo{year}{1995}\natexlab{}.
\newblock \showarticletitle{Taking steps: the influence of a walking technique on presence in virtual reality}.
\newblock \bibinfo{journal}{\emph{ACM Transactions on Computer-Human Interaction}} \bibinfo{volume}{2}, \bibinfo{number}{3} (\bibinfo{year}{1995}), \bibinfo{pages}{201--219}.
\newblock
\href{https://doi.org/10.1145/210079.210084}{doi:\nolinkurl{10.1145/210079.210084}}


\bibitem[Templeman et~al\mbox{.}(1999)]%
        {templeman1999virtualLocomotion}
\bibfield{author}{\bibinfo{person}{James~N. Templeman}, \bibinfo{person}{Patricia~S. Denbrook}, {and} \bibinfo{person}{Linda~E. Sibert}.} \bibinfo{year}{1999}\natexlab{}.
\newblock \showarticletitle{Virtual Locomotion: Walking in Place through Virtual Environments}.
\newblock \bibinfo{journal}{\emph{Presence: Teleoperators and Virtual Environments}} \bibinfo{volume}{8}, \bibinfo{number}{6} (\bibinfo{year}{1999}), \bibinfo{pages}{598--617}.
\newblock
\href{https://doi.org/10.1162/105474699566512}{doi:\nolinkurl{10.1162/105474699566512}}


\bibitem[Usoh et~al\mbox{.}(1999)]%
        {usoh1999walking}
\bibfield{author}{\bibinfo{person}{Martin Usoh}, \bibinfo{person}{Kevin Arthur}, \bibinfo{person}{Mary~C. Whitton}, \bibinfo{person}{Rui Bastos}, \bibinfo{person}{Anthony Steed}, \bibinfo{person}{Mel Slater}, {and} \bibinfo{person}{Frederick~P. Brooks}.} \bibinfo{year}{1999}\natexlab{}.
\newblock \showarticletitle{Walking $>$ walking-in-place $>$ flying, in virtual environments}. In \bibinfo{booktitle}{\emph{Proceedings of the 26th Annual Conference on Computer Graphics and Interactive Techniques}}. \bibinfo{publisher}{Association for Computing Machinery}, \bibinfo{address}{New York, NY, USA}, \bibinfo{pages}{359--364}.
\newblock
\href{https://doi.org/10.1145/311535.311589}{doi:\nolinkurl{10.1145/311535.311589}}


\bibitem[Wang and Lindeman(2012)]%
        {wang2012leaningRevisited}
\bibfield{author}{\bibinfo{person}{Jia Wang} {and} \bibinfo{person}{Rob Lindeman}.} \bibinfo{year}{2012}\natexlab{}.
\newblock \showarticletitle{Leaning-based travel interfaces revisited: frontal versus sidewise stances for flying in {3D} virtual spaces}. In \bibinfo{booktitle}{\emph{Proceedings of the 18th ACM Symposium on Virtual Reality Software and Technology}}. \bibinfo{publisher}{Association for Computing Machinery}, \bibinfo{address}{New York, NY, USA}, \bibinfo{pages}{121--128}.
\newblock
\href{https://doi.org/10.1145/2407336.2407360}{doi:\nolinkurl{10.1145/2407336.2407360}}


\bibitem[Zielasko and Riecke(2021)]%
        {zielasko2021sitOrNot}
\bibfield{author}{\bibinfo{person}{Daniel Zielasko} {and} \bibinfo{person}{Bernhard~E. Riecke}.} \bibinfo{year}{2021}\natexlab{}.
\newblock \showarticletitle{To Sit or Not to Sit in {VR}: Analyzing Influences and (Dis)Advantages of Posture and Embodied Interaction}.
\newblock \bibinfo{journal}{\emph{Computers}} \bibinfo{volume}{10}, \bibinfo{number}{6} (\bibinfo{year}{2021}), \bibinfo{pages}{73}.
\newblock
\href{https://doi.org/10.3390/computers10060073}{doi:\nolinkurl{10.3390/computers10060073}}


\end{thebibliography}

\end{document}